\documentclass[preprintnumbers,amsmath,amssymb]{revtex4}
\usepackage{epsfig}
\usepackage{color}
\usepackage{dsfont}
\usepackage[colorlinks,urlcolor=blue,linkcolor=blue,citecolor=cyan]{hyperref}
\begin{document}
\setlength{\voffset}{1.0cm}
\title{Non-perturbative phase boundaries in the Gross-Neveu model from a stability analysis}
\author{Michael Thies\footnote{michael.thies@gravity.fau.de}}
\affiliation{Institut f\"ur  Theoretische Physik, Universit\"at Erlangen-N\"urnberg, D-91058, Erlangen, Germany}
\date{\today}

\begin{abstract}
Two out of three phase boundaries of the 1+1 dimensional Gross-Neveu model in the chiral limit can be obtained from a standard,
perturbative stability analysis of the homogeneous phases. The third one separating the massive homogeneous phase from the 
kink crystal is non-perturbative and could so far only be inferred from the full solution of the model. We show that this phase
boundary can also be obtained via a modified stability analysis, based on the thermodynamic potential of a single kink or baryon.
The same method works for the massive Gross-Neveu model, so that all phase boundaries of the Gross-Neveu model could
have been predicted quantitatively without prior knowledge of the full crystal solution.
\end{abstract}

\maketitle

\section{Introduction}
\label{sec1}

Originally put forward as an example of a non-gauge theory with asymptotic freedom, the Gross-Neveu (GN) model in 1+1 dimensions \cite{1} has turned out to 
be an inspiring toy model for questions of hot and dense matter as well \cite{2}. The Lagrangian of its simplest version reads
\begin{equation}
{\cal L} = \sum_{i=1}^N \bar{\psi}_i \left(i \gamma^{\mu}\partial_{\mu} -m_0 \right)  \psi_i  + \frac{1}{2}g^2 \left(\sum_{i=1}^N \bar{\psi}_i\psi_i\right)^2.
\label{1.1}
\end{equation}
It describes $N$ species of Dirac fermions interacting via an attractive quartic scalar-scalar interaction. For vanishing bare mass $m_0$, it exhibits
a discrete chiral symmetry ($\psi \to \gamma_5 \psi$, group Z$_2$). We shall only be concerned with the large $N$ limit here ('t Hooft limit \cite{3}, $N \to \infty$
with $Ng^2$ kept constant) where semiclassical computations can be trusted. For fermions, this justifies the use of the relativistic Hartree-Fock (HF) approach.

\begin{figure}
\begin{center}
\epsfig{file=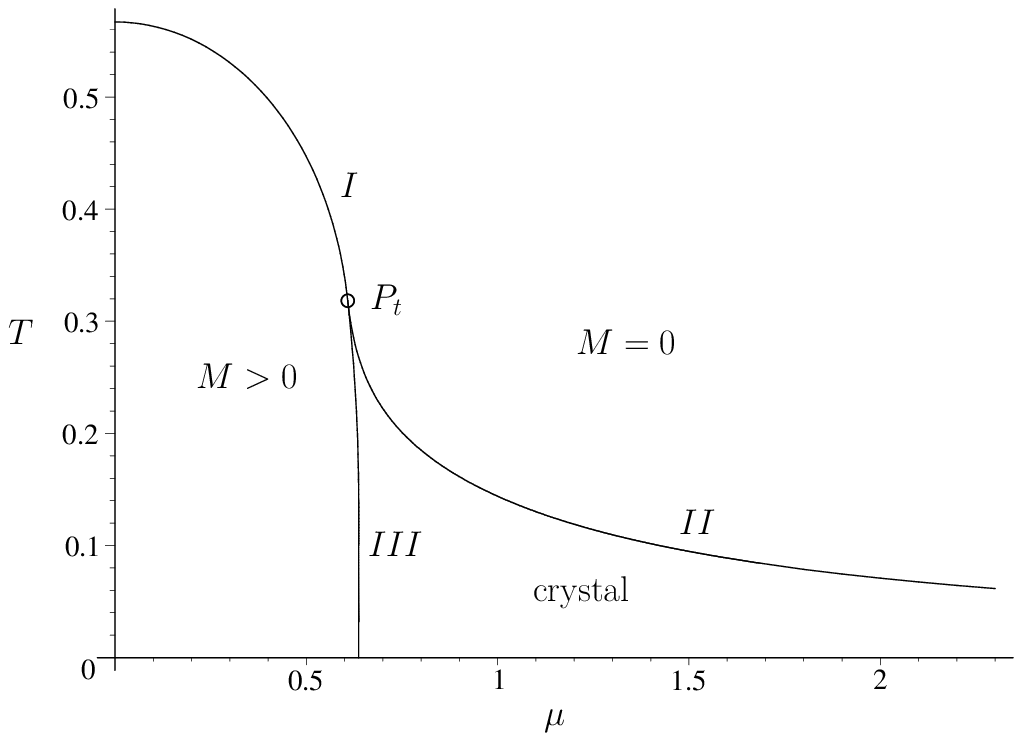,height=8cm,angle=270}
\caption{Phase diagram of the GN model in the chiral limit, featuring three distinct phases separated by 2nd order phase boundaries. The curves   
$I-III$ are discussed in the main text. Adapted from Ref.~\cite{5}.}
\label{fig1}
\end{center}
\end{figure}

\begin{figure}
\begin{center}
\epsfig{file=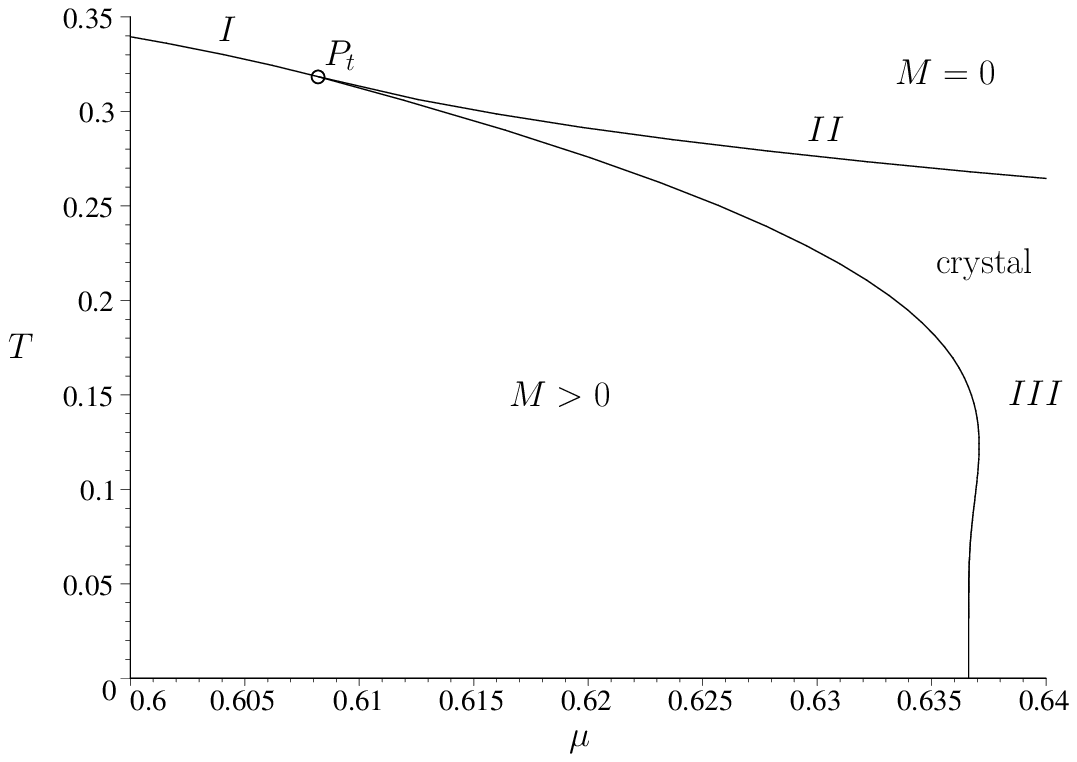,height=8cm,angle=270}
\caption{Zooming in onto the non-perturbative phase boundary $III$ of Fig.~\ref{fig1}, the central topic of the present study.} 
\label{fig2}
\end{center}
\end{figure}

Let us first recall the phase diagram of model (\ref{1.1}) in the chiral limit, see Fig.~\ref{fig1} and Refs.~\cite{4,5}. Here, as in all following figures,
we use units $m=1$ with $m$ the dynamical fermion mass in vacuum. Three distinct phase boundaries meet in a tricritical point $P_{\rm t}$.
Curve $I$ separates the chirally symmetric phase (massless fermions) at high temperatures from a homogeneous broken phase with dynamical mass $M$
at lower temperatures. This phase boundary can be obtained via a perturbative stability analysis, since the fermion mass vanishes continuously 
at the boundary. Curve $II$ separates the symmetric phase from a soliton crystal phase at low temperatures and larger chemical potentials. Since the
amplitude of the periodic modulation of the chiral condensate vanishes continuously along the curve, this boundary is also accessible through
a perturbative stability analysis, the perturbation being a harmonic function rather than a constant mass. In the past, the perturbative phase boundaries
$I$ and $II$ have been rederived by several authors, using canonical perturbation theory \cite{4}, a ``fermion doubler trick"  \cite{5a} or, more recently, the bosonic two-point function 
approach \cite{6}. It is not necessary to know the full solution of the equilibrium problem to pin down these 2nd order phase boundaries. Curve $III$ is the most
intriguing
phase boundary and in the focus of the present work. We show a more detailed plot with blown-up scale of the chemical potential axis in Fig.~\ref{fig2}.
It separates the homogeneous broken phase from the inhomogeneous, kink-antikink crystal phase. While it is also a 2nd order phase boundary, the
instability here is with respect to formation of a single kink rather than ripples. A kink may be thought of as the one-dimensional analogue of a 
domain wall, separating regions with homogeneous condensates of opposite signs. 
Since the kink amplitude is determined by the thermal mass $M$, it does
not vanish at the phase boundary, but jumps discontinuously. Yet, since the perturbation is localized in space, all bulk thermodynamic observables are
continuous across the phase boundary. This behaviour does not really fit into the common use of the stability analysis. This is presumably the reason
why curve $III$ has not yet been reproduced by any calculation independent of the full, thermal Hartree-Fock approach (or a numerical simulation on the 
lattice) to date, to the best of our knowledge. It is the purpose of the present paper to show how one can nevertheless reproduce this kind of non-perturbative,
2nd order phase boundary by a modified stability analysis, provided one looks for localized rather than periodically oscillating perturbations
of the homogeneous mean field and avoids using perturbation theory. Needless to say, this can only be applied to continuous (2nd order) phase
transitions. Thus the present method would not work for the massive chiral Gross-Neveu model, where the phase
transition analogous to curve $III$ has been found numerically to be of first order \cite{7}. 

\begin{figure}
\begin{center}
\epsfig{file=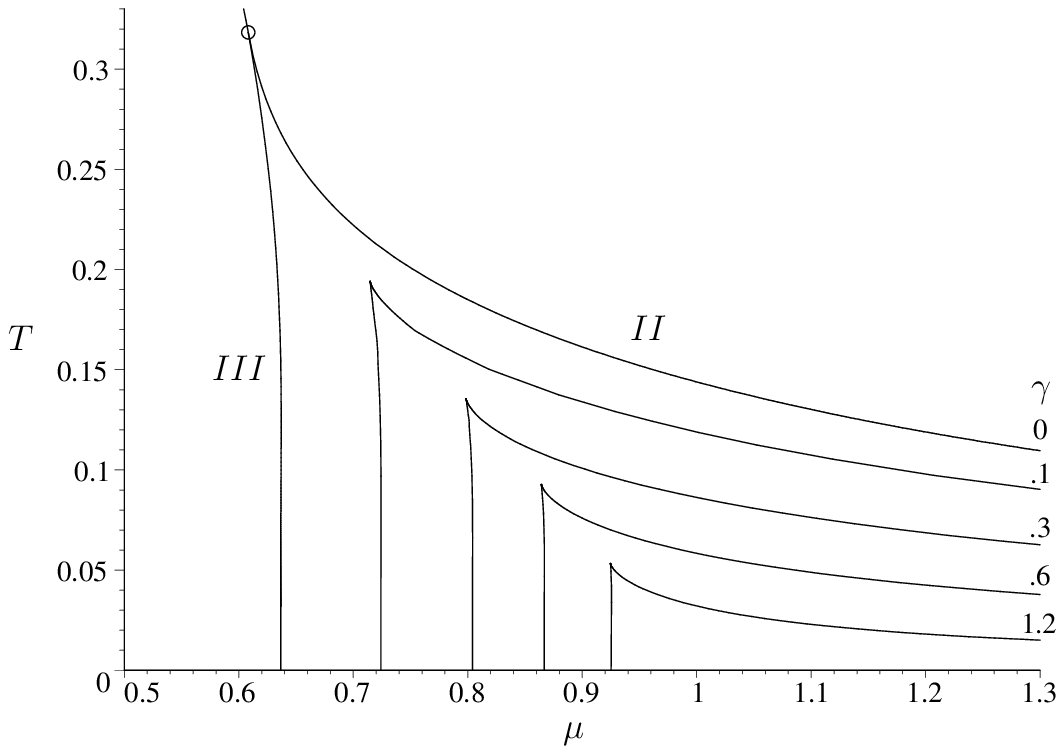,height=8cm,angle=270}
\caption{Comparison of the phase diagrams of the massless (outmost curve, $\gamma=0$) and massive ($\gamma=.1,.3,.6,1.2$ from top to bottom)
GN models. The confinement parameter $\gamma$ is a measure for the bare mass defined in Eq.~(\ref{3.2}) of Sect.~\ref{sec3} below. Adapted from Ref.~\cite{8}.}
\label{fig3}
\end{center}
\end{figure}

What changes if we turn on the bare fermion mass $m_0$, breaking the discrete chiral symmetry explicitly? Clearly, the symmetric phase with massless
fermions must disappear and only a homogeneous and an inhomogeneous phases remain. Consequently, phase boundary $I$ disappears in favor of
a cross-over. Phase boundaries $II$ and $III$ survive, ending in a cusp at a tricritical point, see Fig.~\ref{fig3} and Refs.~\cite{2,8}. Curve $II$ can again be
obtained from a conventional stability analysis, except that now the unperturbed mean field is constant rather than 0. Along curve $III$, instability
with respect to creation of a kink-antikink baryon of the massive Gross-Neveu model occurs.

It should be mentioned that the $T=0$ base points of curves $III$ are always located at $\mu=M_{\rm B}/N$ with $M_{\rm B}$ the baryon mass of the 
corresponding massive GN model. The reason is the fact that the grand canonical potential for a baryon with maximal fermion number $N_f=N$ at $T=0$ 
is given by
\begin{equation}
\Psi = M_{\rm B}-\mu N
\label{1.2}
\end{equation}
and vanishes at that value of $\mu$. Hence, at least one point of the phase boundaries $III$ can be found exactly without reference to the full HF calculation
for dense matter, namely by computing the baryon mass. This makes it tempting to look for a generalization of such a shortcut to finite temperature,
the aim of the present paper.

This paper is organized as follows. Sect.~\ref{sec2} is dedicated to the chiral limit. In Sect.~\ref{sec2a}, the thermodynamic potential of a single baryon
in contact with a heat bath and a fermion reservoir is set up. In Sect.~\ref{sec2b} we discuss numerical results based on this formalism. In Sect.~\ref{sec2ba}
we show the close relationship between the phase diagram of a single baryon and that of the infinite system, notably concerning the non-perturbative
phase boundary of the latter. In Sect.~\ref{sec2bb}, further results are presented characterizing the baryon at finite temperature and chemical potential.
Sect.~\ref{sec3} addresses the massive GN model. In Sect.~\ref{sec3a} we generalize the results of Sect.~\ref{sec2a} to finite bare fermion masses.
Sect.~\ref{sec3b} shows numerical results, first for the phase boundary in Sect.~\ref{sec3ba} and then for thermodynamic observables in Sect.~\ref{sec3bb}.
In Sec.~\ref{sec4} we confirm our results by rederiving them starting from the known, full solution of the GN model thermodynamics. This is followed by
a short summary and outlook in Sect.~\ref{sec5}.

\section{Gross-Neveu model in the chiral limit}
\label{sec2}

\subsection{Single baryon at finite temperature and chemical potential}
\label{sec2a}

Baryons of the GN model in the chiral limit are known since the work of Dashen, Hasslacher and Neveu (DHN) \cite{9} and have been further explored by
several authors \cite{10,11,12}. For an elementary approach using the relativistic Hartree-Fock (HF) method, see \cite{13}. In the large $N$ limit, the full solution
is available in closed analytical form. The scalar mean field has the kink-antikink shape 
\begin{equation}
S(x)=m+ym\left[\tanh(ymx-c_0)-\tanh(ymx+c_0)\right], \quad c_0 = \frac{1}{2} {\rm artanh\,} y ,
\label{2.1}
\end{equation}  
with $m$ the dynamical fermion mass in vacuum and $y\in [0,1]$ a parameter governing its precise form and fermion number. All single particle wave
functions and eigenvalues are known exactly. The spectrum of continuum states ($\pm E_q =\pm \sqrt{q^2+m^2}$) and bound states ($\pm E_0 = \pm m\sqrt{1-y^2}$)
is symmetric about 0, reflecting charge conjugation symmetry. The fermion-baryon phaseshifts are given by
\begin{equation}
\delta(q)= \arctan \frac{2ymq}{q^2-y^2m^2}.
\label{2.2}
\end{equation}
The parameter $y$ can be related to the fermion number by means of the self-consistency condition
\begin{equation}
\arctan \frac{\sqrt{1-y^2}}{y} =\frac{\pi}{2}\left(1-\frac{n}{N}\right).
\label{2.3}
\end{equation}
Here, $n$ is the number of fermions in the positive energy bound state, the negative energy bound state and all negative energy continuum states 
being fully occupied. It is at the same time the fermion number of the baryon. The mass of the baryon is found to be 
\begin{equation}
\frac{M_{\rm B}}{N} = \frac{2ym}{\pi}.
\label{2.4}
\end{equation}
For maximum fermion number $n=N$, $y$ approaches 1 and the kink and antikink are infinitely separated. They decouple
into a pair of independent kink and antikink (sometimes referred to as CCGZ-kink \cite{14}) with profile
\begin{equation}
S_{\rm kink}(x) = \pm m \tanh(mx)  .
\label{2.5}
\end{equation}
The 2nd ingredient that we shall need is the GN model with a constant scalar condensate, but at finite temperature and chemical potential. This is the 
homogeneous ordered phase with broken chiral symmetry, stable in a certain region of the phase diagram. It has first been studied by Wolff \cite{15}. 
The language we shall use here is that of thermal HF, tailored to equilibrium problems in the semiclassical regime. The system generates dynamically 
a thermal fermion mass $M$ depending on temperature $T$ and chemical potential $\mu$ through a self-consistency condition.
We recall the renormalized grand potential density ($\beta=1/T, E_q=\sqrt{M^2+q^2}$)
\begin{equation}
\frac{\psi}{N}  = \frac{M^2}{2\pi} \left( \ln \frac{M}{m} - \frac{1}{2} \right) - \frac{1}{\beta \pi} \int_0^{\infty} dq \ln\left[ \left( 1+e^{-\beta(E_q-\mu)}  \right) \left(   1+e^{-\beta(E_q+\mu)}   \right) \right].
\label{2.6}
\end{equation}
Depending on $T$ and $\mu$, its minimum is either at $M=0$ (symmetric phase), or it is a solution of the equation
\begin{equation}
\ln \frac{M}{m} + \int_0^{\infty} dq \frac{1}{E_q} \left( \frac{1}{e^{\beta(E_q-\mu)}+1}+ \frac{1}{e^{\beta(E_q+\mu)}+1} \right) = 0.
\label{2.7}
\end{equation}
In case there are several solutions, one has to choose the absolute minimum of the grand potential density (\ref{2.6}).

So far, this is all well known. We now try something new, namely to study the thermodynamics of a single DHN baryon at finite temperature and chemical potential.
A priori, it is not clear whether such a system exists and is thermodynamically stable. We first have to guess an ansatz for the scalar mean field.
We simply assume that the functional form of $S(x)$ is the same as for a baryon in the vacuum, except that the parameter $m$ has to be replaced by the thermal
mass $M$ to match the homogeneous medium asymptotically. The parameter $y$ will be left open for the moment. The main advantage of this ansatz is the fact that we 
can solve the Dirac-HF equation exactly, using results from the DHN baryon. Notice that this ansatz fits into the idea of a stability analysis, since we
can decompose the mean field as
\begin{eqnarray}
S(x) & = & M + \Delta S(x),
\nonumber     \\
\Delta S(x) & = & yM\left[\tanh(yMx-c_0)-\tanh(yMx+c_0)\right], \quad c_0 = \frac{1}{2} {\rm artanh} y.
\label{2.8}
\end{eqnarray}
In standard applications of the stability analysis, the correction term $\Delta S(x)$ can be treated perturbatively since its amplitude vanishes at the 
phase boundary. Here, this is not the case, the amplitude of $\Delta S(x)$ being determined by the thermal mass $M$. 
The simple ansatz (\ref{2.8}) enables us to proceed nevertheless, namely by solving the Dirac-HF equation exactly.
We would like to compute the grand canonical potential of a single baryon in contact with a heat bath and a particle reservoir.
In contrast to Eq.~(\ref{2.6}) for the infinite medium, we are not interested in the grand potential density, but in the grand canonical potential itself.
As is familiar from the HF approach, it consists of a single particle term and a double counting correction for the interaction energy,
\begin{eqnarray}
\Psi & = & \Psi_{\rm sp}+ \Psi_{\rm dc},
\nonumber\\
\frac{\Psi_{\rm sp}}{N} & = &   -\frac{1}{\beta} \sum_n   \ln \left[ \left( 1+e^{-\beta(E_n-\mu)} \right)  \left(1+e^{\beta(E_n+\mu)} \right)    \right],
\nonumber \\
\frac{\Psi_{\rm dc}}{N} & = & \frac{1}{2 Ng^2}\int dx S^2(x).
\label{2.9}
\end{eqnarray}
Let us assume temporarily that we work in a finite box of length $L$, so that the spectrum gets discretized. Since we have pulled out the factor $N$
and spelled out positive and negative eigenstates in $\Psi_{\rm sp}$, the sum over $n$ runs only over positive energy single particle orbits and does not include flavor.
In the spirit of a stability analysis, we shall compute everything relative to the homogeneous thermal ground state. For the double counter, this
amounts to a trivial subtraction
\begin{equation}
\frac{\Psi_{\rm dc}}{N} \to \frac{1}{2Ng^2} \int dx \left( S^2(x)-M^2\right) = - \frac{2yM}{\pi} \ln \frac{\Lambda}{m}.
\label{2.10}
\end{equation}
The integral is elementary \cite{13}, and we have eliminated the coupling constant with the help of the 
vacuum gap equation
\begin{equation}
\frac{\pi}{Ng^2} = \ln \frac{\Lambda}{m}.
\label{2.11}
\end{equation}
Turning to the single particle contribution, consider the unbound states first. 
As usual we split the single particle contribution up into UV divergent and convergent pieces, 
\begin{eqnarray}
\Psi_{\rm sp} & = & \Psi_{\rm sp}^{(1)} + \Psi_{\rm sp}^{(2)},
\nonumber \\
\frac{\Psi_{\rm sp}^{(1)}}{N} & = & - \sum_n (E_n+\mu),
\nonumber \\
\frac{\Psi_{\rm sp}^{(2)}}{N} & = &   -\frac{1}{\beta} \sum_n \ln \left[\left(1+e^{-\beta(E_n-\mu)}\right) \left(1+e^{-\beta(E_n+\mu)}\right)    \right].
\label{2.12}
\end{eqnarray}
Closely following the calculation of the baryon mass \cite{13}, we evaluate the difference between baryon and medium 
by carefully counting the modes (the baryon has discrete bound states, the medium only continuum states \cite{16}). 
Assuming periodic boundary conditions in a box of length $L$, the fermion momenta in the baryon and the medium 
are related by
\begin{equation}
q_n = q_n^0 - \frac{1}{L} \delta(q_n^0),  \quad q_n^0 = \frac{2\pi}{L} n,
\label{2.13}
\end{equation}
with the phase shift (\ref{2.2}) (replacing $m$ by $M$). We denote the subtracted divergent contribution in (\ref{2.12}) by
\begin{equation}
\frac{\Psi_{\rm sp}^{(1)}}{N} = - \sum_n (E_n+\mu) \to \Delta_E+ \Delta_{\mu}.
\label{2.14}
\end{equation}
The subtracted sum over the (negative) single particle energies becomes
\begin{eqnarray}
\Delta_E & = & - \sum_n \left[ E(q_n) - E(q_n^0) \right]
\nonumber \\
& = & \sum_n \frac{\delta(q_n^0)}{L} \left. \frac{d E_q}{dq} \right|_{q_n^0}.
\label{2.15}
\end{eqnarray}
Taking the thermodynamic limit $L \to \infty$, we convert the sum into an integral and regularize it with the same cutoff
that has been used in the gap equation
\begin{eqnarray}
\Delta_E & = & \int_{-\Lambda/2}^{\Lambda/2} \frac{dq}{2\pi}  \delta(q) \frac{d E_q}{dq}
\nonumber \\
& = & \frac{2 y M}{\pi} + \int_{-\Lambda/2}^{\Lambda/2}  \frac{dq}{2 \pi} \frac{2yM}{q^2+y^2M^2} E_q.
\label{2.16}
\end{eqnarray}
In the integral, the standard relation between phase shift and density of states is recognized
\begin{equation}
\rho(q) = \frac{d}{dq} \delta(q) = - \frac{2yM}{q^2+y^2M^2}.
\label{2.17}
\end{equation} 
The first term in (\ref{2.16}) arises from a partial integration or, alternatively, from a modification of the cutoff when
going from $q_n^0$ to $q_n$; it is a kind of anomalous term.
The result of the integration in (\ref{2.16}) can be taken over literally from the baryon mass calculation,
\begin{equation}
\Delta_E  =   \frac{2yM}{\pi} \left( 1 + \ln \frac{\Lambda}{M}+ \frac{\sqrt{1-y^2}}{y} \arctan \frac{\sqrt{1-y^2}}{y}\right).
\label{2.18}
\end{equation}
For the chemical potential part of $\Psi_{\rm sp}^{(1)}$, we may now use the density of states (\ref{2.17})
to replace the discrete sum by an integral,
\begin{equation}
\Delta_{\mu} =    \mu \int_{-\infty}^{\infty} \frac{dq}{2 \pi} \frac{2yM}{q^2+y^2M^2} = \mu.
\label{2.19}
\end{equation}
There is no anomalous term since the integral is convergent. The result just reflects the fact that the interacting
theory has one (negative energy) continuum state less than the free theory, due to the bound state of the potential.
This completes the calculation of the divergent single particle part. Along similar lines, the convergent part and 2nd
term in Eq.~(\ref{2.12}) yields upon subtraction 
\begin{equation}
\frac{\Psi_{\rm sp}^{(2)}}{N}  = \frac{1}{\pi \beta} \int_{0}^{\infty} dq \frac{2yM}{q^2+y^2M^2} \ln \left[ \left( 1+e^{-\beta(E_q-\mu)} \right) \left( 1+e^{-\beta(E_q+\mu)} \right) \right].
\label{2.20}
\end{equation} 
Finally, we have to add the contribution from the pair of bound states, 
\begin{eqnarray}
\frac{\Psi_{\rm sp}^{(0)}}{N}  =  - \frac{1}{\beta} \ln \left[ \left( 1+e^{-\beta(E_0-\mu)} \right) \left( 1 + e^{\beta(E_0+\mu)}\right) \right]   .    
\label{2.21}
\end{eqnarray}
Collecting all terms yields the following expression for the grand canonical potential per color of a single baryon
\begin{eqnarray}
\frac{\Psi}{N} & = &  \frac{2yM}{\pi} \left( 1 + \ln \frac{m}{M}+ \frac{\sqrt{1-y^2}}{y} \arctan \frac{\sqrt{1-y^2}}{y}\right)
\nonumber \\
& &  + \mu  - \frac{1}{\beta} \ln \left[ \left( 1+e^{-\beta(E_0-\mu)} \right) \left( 1 + e^{\beta(E_0+\mu)}\right) \right]
\nonumber  \\
& &  + \frac{1}{\pi \beta} \int_{0}^{\infty} dq \frac{2yM}{q^2+y^2M^2} \ln \left[ \left( 1+e^{-\beta(E_q-\mu)} \right) \left( 1+e^{-\beta(E_q+\mu)} \right) \right].
\label{2.22}
\end{eqnarray}
Notice that the cutoff dependence has disappeared owing to a cancellation of the $\ln \Lambda$-terms between the double counting correction and the divergent single particle part.
In contrast to the DHN baryon in vacuum, here the parameter $y$ cannot be related to the occupation fraction of the bound state. We cannot account for partial occupation of the 
discrete state when working with a chemical potential. We therefore consider $y$ as a variational parameter and use it 
to minimize the grand potential. It actually can serve as an order parameter for breaking of translational symmetry, since $y=0$ 
corresponds to the homogeneous condensate and $0<y\le 1$ to a spatially localized inhomogeneity.

\subsection{Numerical results}
\label{sec2b}

\subsubsection{Phase transition}
\label{sec2ba}

\begin{figure}
\begin{center}
\epsfig{file=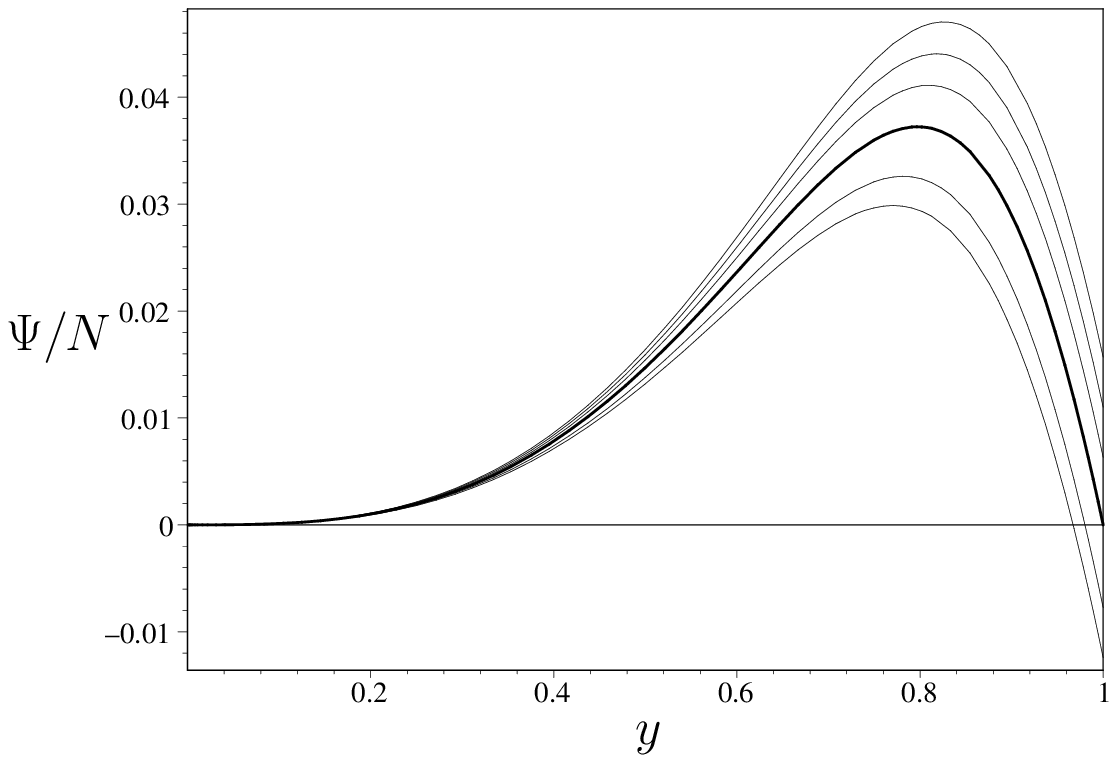,height=8cm,angle=270}
\caption{Thermodynamic potential of a single baryon in the chiral limit, Eq.~(\ref{2.22}), as a function of $y\in [0,1]$. The curves correspond to $T=0.15$ and 
$\mu=.62,.625,.63,\mu_c,.645,.65$ from top to bottom and show a first order phase transition at $\mu_c=.6367374$. Here, $y$ jumps from 0 to 1.} 
\label{fig4}
\end{center}
\end{figure}

This part of our study is guided by the phase diagram of the GN model in the chiral limit, Fig.~\ref{fig1}. The homogeneous broken phase is stable in the region
to the left of phase boundaries $I$ and $III$. We are interested in the non-perturbative phase boundary $III$, see also Fig.~\ref{fig2} for more details. Therefore
we explore the behavior of the 
grand canonical potential (\ref{2.22}) as we cross the phase boundary $III$ at constant temperature. All we have to do is choose ($\mu,T$), determine the 
thermal mass via Eq.~(\ref{2.7}) and insert these parameters into Eq.~(\ref{2.22}).
By way of example, take $T=0.15$ and evaluate $\Psi$ for 6 different values of $\mu$ in the vicinity of the 
phase boundary at $\mu_c=0.6367374$. Since we have not yet determined the parameter $y$, we show $\Psi$ as a function of $y$ (between 0 and 1) for
these 6 points in Fig.~\ref{fig4}. Always taking the value of $y$ where $\Psi$ is minimal, we find a 1st order phase transition at $\mu=\mu_c$ 
where $y$ jumps from 0 to 1 and translation invariance breaks down. Notice that the minimum at $y=1$ could not have been found by taking the derivative 
of $\Psi$ with respect to $y$. It coincides with the endpoint of the interval [0,1] of allowed values of $y$. 

If one repeats this calculation at different temperatures (below the tricritical point), one always finds the same behavior: Only the values $y=0$ and $y=1$
are allowed, with a first order phase transition in between. The value $y=1$ means that kink and antikink are infinitely separated and have the simple 
profile from Eq.~(\ref{2.5}). For $y=1$, our result (\ref{2.22}) for the grand canonical potential simplifies as follows,
\begin{eqnarray}
\frac{\Psi}{N} & = &  \frac{2M}{\pi} \left( 1+ \ln \frac{m}{M}\right) +\mu   - \frac{2}{\beta} \ln  \left( 1+e^{\beta \mu} \right) 
\nonumber \\
& &  + \frac{2M}{\pi \beta} \int_{0}^{\infty} dq \frac{1}{E_q^2} \ln \left[ \left( 1+e^{-\beta(E_q-\mu)} \right) \left( 1+e^{-\beta(E_q+\mu)} \right) \right].
\label{2.23}
\end{eqnarray}
If we divide $\Psi/N$ by 2, the result can be interpreted as grand canonical potential of a single kink or domain wall.
Thus, the phase transition can also be viewed as instability of the homogeneous system with respect to (single) kink formation.

\begin{figure}
\begin{center}
\epsfig{file=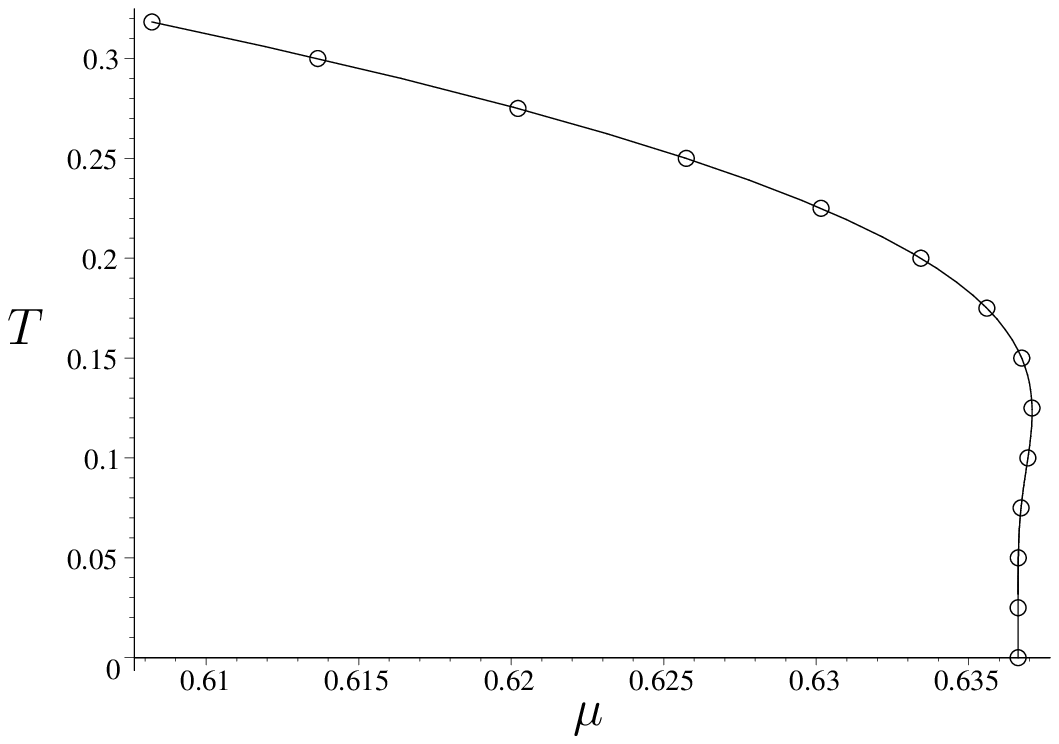,height=8cm,angle=270}
\caption{Comparison of 1st order phase boundary of a single baryon (circles, this work) with 2nd order phase boundary $III$ of the chiral GN model (line, 
Refs.~\cite{4,5}).}
\label{fig5}
\end{center}
\end{figure}

What is the relationship between this first order phase boundary of the single baryon or kink system and the 2nd order phase boundary of the GN model,
curve $III$, shown in Figs.~\ref{fig2},\ref{fig3}? In Fig.~\ref{fig5}, we compare these two phase boundaries coming from independent sources. We choose
the blown-up scale on the $\mu$-axis for a more quantitative comparison. The agreement is perfect, supporting our ansatz for the baryon at finite
temperature and chemical potential.
Had we chosen another ansatz, for instance by prescribing the value of $y$ arbitrarily rather than searching for the minimum of $\Psi$, our calculation
would be of variational type and yield a (spurious) phase boundary inside the crystal region, to the right of the true one. Thus we are able to
reproduce the non-perturbative phase transition without full HF calculation. Only one-baryon input enters the calculation and we do not need to know
anything about the crystal structure of the inhomogeneous phase. In this sense, we are now in a similar position as when  
analyzing phase boundaries $I$ and $II$ using the standard, perturbative stability analysis. This also sheds some light on the question why
a non-perturbative phase transition can still be a continuous, 2nd order one. A single kink or baryon in an infinite medium cannot induce
a discontinuity in any thermodynamic observable.

Now suppose that we would have performed the same calculation, but that the full phase diagram of the GN model would not have been known.
Is there a way of telling whether our ansatz for the condensate was the correct one? This is indeed the case. What one would have to do is
check the self-consistency of the HF solution, as it has been done for the original DHN baryon in vacuum. The (thermal) self-consistency condition reads
\begin{eqnarray}
S(x) & = &  - Ng^2 \langle \bar{\psi}\psi \rangle_{\rm th}
\nonumber \\
& = & - N g^2 \int_0^{\Lambda/2}\frac{dq}{\pi} \bar{\psi}_q^{(+)}\psi_q^{(+)} \left( \frac{1}{1+ e^{\beta(E_q-\mu)}} + \frac{1}{1+e^{\beta(E_q+\mu)}} -1 \right).
\label{2.24}
\end{eqnarray}
For $y=1$, the bound states do not contribute, while the continuum states yield the scalar density \cite{13} 
\begin{equation}
\bar{\psi}_q^{(+)} \psi_q^{(+)} = \frac{S(x)}{E_q}.
\label{2.25}
\end{equation}
Upon using the gap equation and performing the integral over $1/E_q$,
\begin{equation}
\frac{\pi}{Ng^2} = \ln \frac{\Lambda}{m}, \quad \int \frac{dq}{2\pi} \frac{1}{E_q} = \frac{1}{\pi} \ln \frac{\Lambda}{M},
\label{2.26}
\end{equation}
one finds that Eq.~(\ref{2.24}) is reduced to the equation for the thermal mass, Eq.~(\ref{2.7}). Thus, provided we use the correct thermal mass $M(\mu,T)$, the 
ansatz (\ref{2.8}) with $y=1$ is indeed self-consistent. 

\subsubsection{Thermodynamic observables}
\label{sec2bb}

\begin{figure}
\begin{center}
\epsfig{file=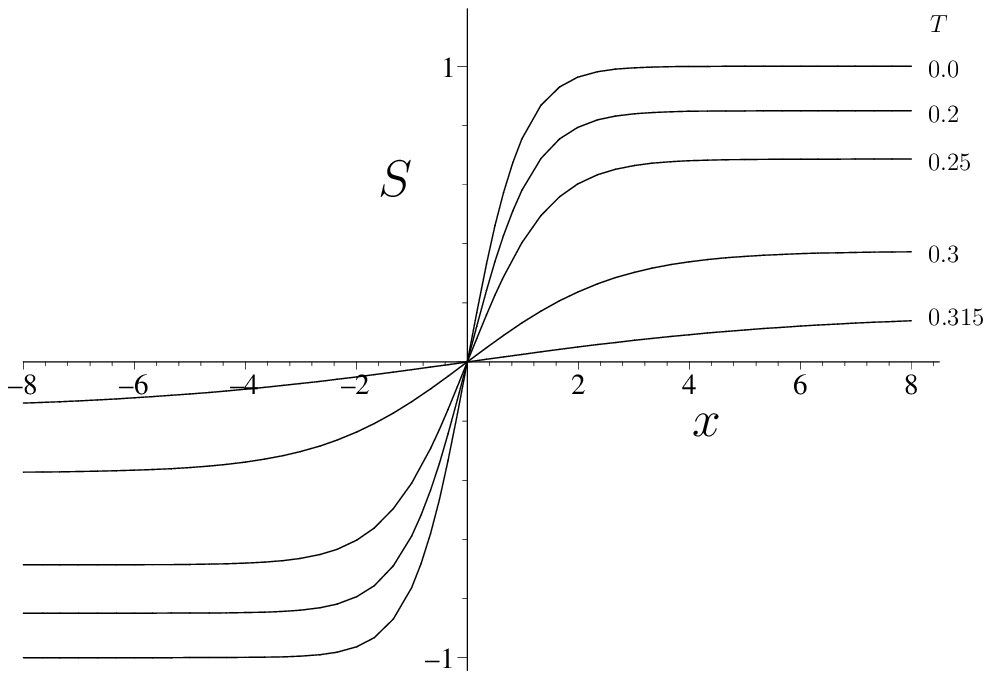,height=8cm,angle=270}
\caption{Illustration of melting of the kink with rising temperature, as one moves up along the phase boundary. At the tricritical point ($\mu_t=.608221,T_t=.318329$), both 
the thermal mass and the kink disappear.} 
\label{fig6}
\end{center}
\end{figure}

\begin{figure}
\begin{center}
\epsfig{file=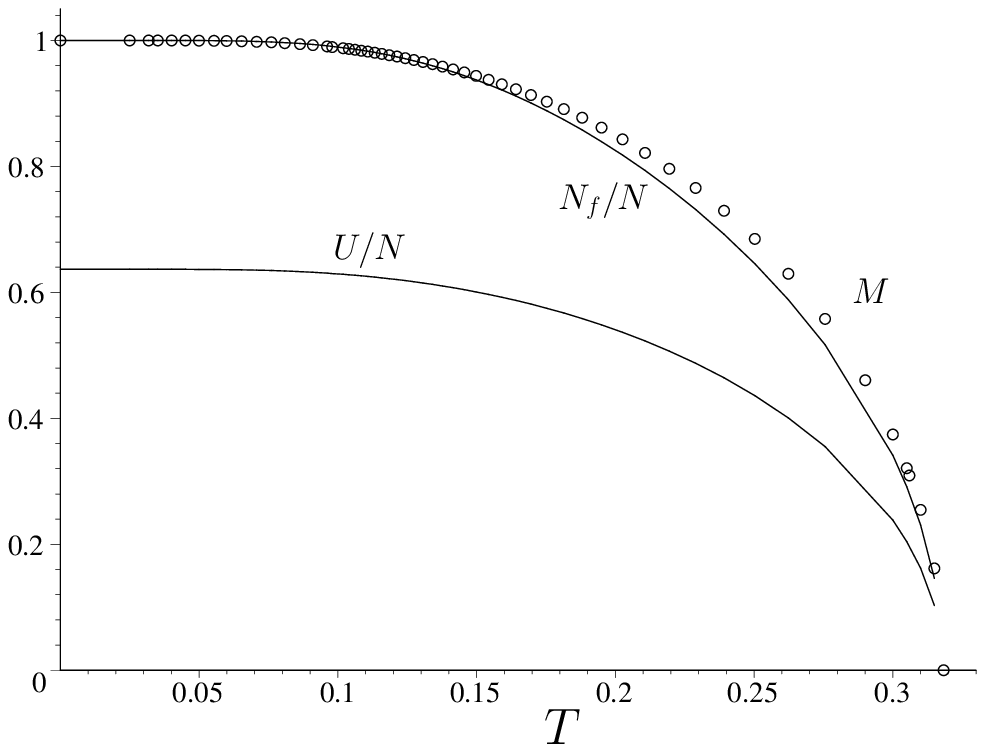,height=8cm,angle=270}
\caption{Thermodynamic observables of a single baryon along the phase boundary, plotted against temperature. Curves: Fermion number $N_f$ and internal
energy $U$ per flavor. Circles: Thermal masses $M$, for comparison. The temperature dependence of the observables is strongly correlated with that of 
the fermion mass.}
\label{fig7}
\end{center}
\end{figure}

\begin{figure}
\begin{center}
\epsfig{file=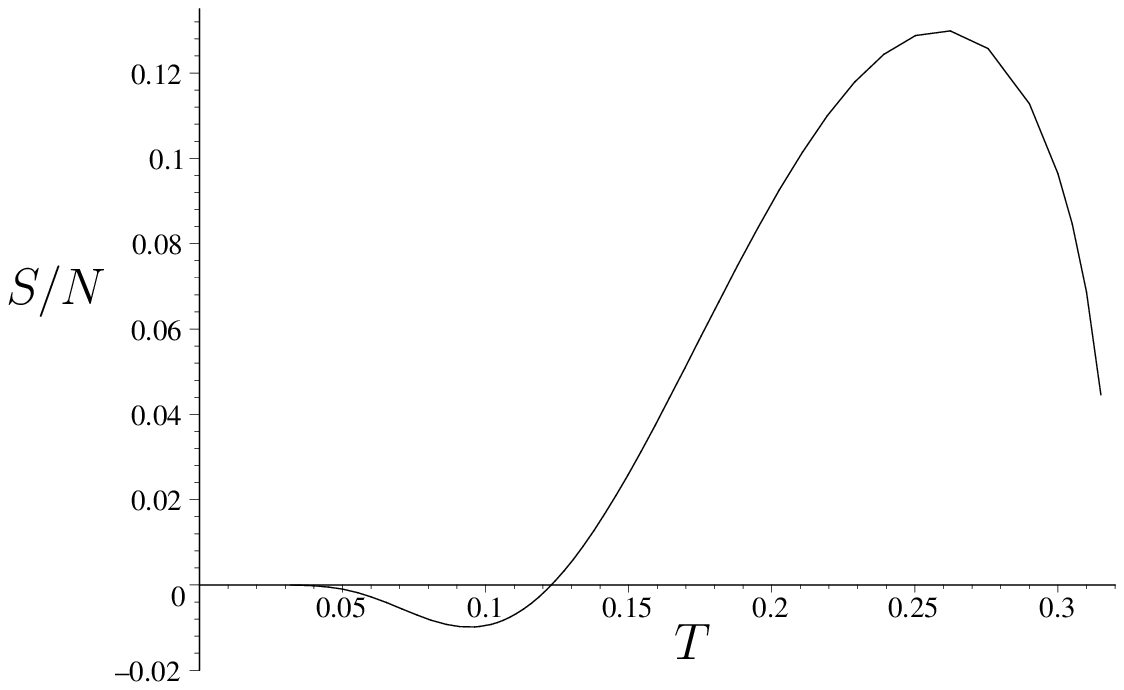,height=8cm,angle=270}
\caption{Entropy of a single baryon along the phase boundary, plotted vs. temperature. The structure is due to a competition between bound state 
and continuum contributions.} 
\label{fig8}
\end{center}
\end{figure}

We have shown how one can create a single, stable kink at a finite temperature, namely by fine-tuning the chemical potential to curve $III$ of 
the GN phase diagram. Since this object is kind of new, it is interesting to ask about its physical properties. 
But let us first illustrate the melting of the kink with rising temperature. The shape of the kink only depends on the scale $M$, see Eq.~(\ref{2.5}). 
Since $M$ decreases with increasing temperature, the kink becomes softer and softer until it disappears at the tricritical temperature.
This is shown in Fig.~\ref{fig6} for a few values of the temperature. Other properties must be phrased in terms of thermodynamic observables that exist in the grand canonical ensemble.
They can be computed along the phase boundary using derivatives of $\Psi$ and comprise notably the fermion number
\begin{equation}
N_f = - \frac{\partial \Psi}{\partial \mu},
\label{2.27}
\end{equation}
the entropy
\begin{equation}
S = - \frac{\partial \Psi}{\partial T},
\label{2.28}
\end{equation}
and the internal energy
\begin{equation}
U = TS+\mu N_f + \Psi.
\label{2.29}
\end{equation} 
In the last equation, we could have dropped $\Psi$ which vanishes along the relevant phase boundary.  
Since the chemical potential varies only by a few percent along the phase boundary, we plot the observables against temperature, always
computing them on the phase boundary where the single baryon system is stable. Fig.~\ref{fig7} shows the fermion number and the internal energy per flavor. 
The temperature dependence directly reflects the behavior of the thermal mass, also shown in Fig.~\ref{fig7}. The starting points at $T=0$ of all 3 curves
have a simple interpretation: Fermion number is maximal ($N_f/N=1$), indicating a fully occupied bound state at $T=0$. The internal energy
matches the corresponding baryon mass ($M_{\rm B}/N=2/\pi$). The fermion mass reaches $m=1$ by our choice of units in the figures.
A new observable without correspondence at $T=0$ is the entropy shown in Fig.~\ref{fig8}. It starts out negative and turns positive at higher temperatures,
apparently as a result of the competition between bound state ($S>0$) and continuum ($S<0$) contributions.

\section{Massive Gross-Neveu model}
\label{sec3}

\subsection{Single baryon at finite temperature and chemical potential}
\label{sec3a}

In the massive GN model, we keep the bare mass $m_0$ in the Lagrangian (\ref{1.1}). The vacuum gap equation (\ref{2.11})
has to be replaced by
\begin{equation}
\frac{\pi}{Ng^2} = \gamma + \ln \frac{\Lambda}{m}
\label{3.1}
\end{equation}
with the confinement parameter 
\begin{equation}
\gamma = \frac{m_0}{m} \ln \frac{\Lambda}{m}.
\label{3.2}
\end{equation}
Like the physical fermion mass $m$, $\gamma$ is renormalization group invariant and owes its name to the confinement of kinks, not of fermions.
(Note the discrepancy in our definition of $\gamma$ with that used in Refs.~\cite{17,18} by a factor of $\pi$). In the vacuum, the 
mean field of the baryon has the same functional form as in the chiral limit, but the self-consistency condition (\ref{2.3}) changes into \cite{17,18}
\begin{equation}
\arctan \frac{\sqrt{1-y^2}}{y}  = \frac{\pi}{2} \left( 1- \frac{n}{N} \right) + \gamma  \frac{y}{\sqrt{1-y^2}}.
\label{3.3}
\end{equation}
The baryon mass is given by
\begin{equation}
\frac{M_{\rm B}}{N} = \frac{2m}{\pi}\left( y  + \gamma {\rm \,artanh\,} y\right),
\label{3.4}
\end{equation}
generalizing Eq.~(\ref{2.4}). What changes in the thermodynamic calculations? Both in the homogeneous medium and in the single baryon cases, $\gamma$
enters the thermal HF calculation only through the double counting correction. The single particle contributions can be taken over literally. 
For the homogeneous Fermi gas,
\begin{eqnarray}
\frac{(M-m_0)^2}{2Ng^2} & = &  \frac{1}{2\pi}\left(\gamma + \ln \frac{\Lambda}{m}\right) (M^2-2Mm_0)
\nonumber \\
& = & \frac{M^2}{2\pi} \ln \frac{\Lambda}{m} + \frac{\gamma M}{2\pi} (M-2m).
\label{3.5}
\end{eqnarray}
Since the $\gamma$ independent term is already included in expression (\ref{2.6}),
the grand potential density for the massive GN model can be simply obtained as 
\begin{equation}
\left. \frac{\psi}{N} \right|_{\gamma} = \left. \frac{\psi}{N} \right|_{0} +\frac{\gamma M}{2\pi} \left( M-2m \right) 
\label{3.6}
\end{equation}
where the $\gamma=0$ term stands for expression (\ref{2.6}). Minimizing with respect to the thermal mass $M$, we 
see immediately that the symmetric solution $M=0$ does no longer exist. The mass equation (\ref{2.7}) gets replaced by
\begin{equation}
\ln \frac{M}{m} + \int_0^{\infty} dq \frac{1}{Eq} \left( \frac{1}{e^{\beta(E_q-\mu)}+1} + \frac{1}{e^{\beta(E_q+\mu)}+1}\right) = \gamma \left( \frac{m}{M}-1 \right)
\label{3.7}
\end{equation}
The solution for $M$ with the minimum value of $\psi/N$ yields the thermal mass as a function of ($\mu,T,\gamma$). 
Likewise, in the thermal HF calculation for a single baryon,
the (subtracted) double counting correction $\Psi_{\rm dc}$ of Eq.~(\ref{2.9}) is modified into 
\begin{eqnarray}
\frac{\Psi_{\rm dc}}{N} & \to & \frac{1}{2Ng^2} \int dx \left[ (S-m_0)^2-(M-m_0)^2 \right]
\nonumber \\
& = & \frac{1}{2\pi} \ln  \frac{\Lambda}{m}  \int dx \left(S^2-M^2 \right) + \frac{2 \gamma}{\pi} \left( m {\rm \,artanh\,}y - yM  \right).
\label{3.8}
\end{eqnarray}
Only the last term proportional to $\gamma$ is new and has to be added to the right hand side of Eq. (\ref{2.22}),
\begin{equation}
\left. \frac{\Psi}{N} \right|_{\gamma} = \left. \frac{\Psi}{N} \right|_{0} +  \frac{2 \gamma}{\pi} \left( m {\rm \,artanh\,}y - yM  \right) .
\label{3.9}
\end{equation}
Otherwise, the calculation is the same as in the chiral limit. As explained above, the $y$ parameter cannot be related to fermion number in the grand
canonical ensemble. Therefore we again have to determine $y$ by minimization of the thermodynamic potential.

\subsection{Numerical results}
\label{sec3b}

\begin{figure}
\begin{center}
\epsfig{file=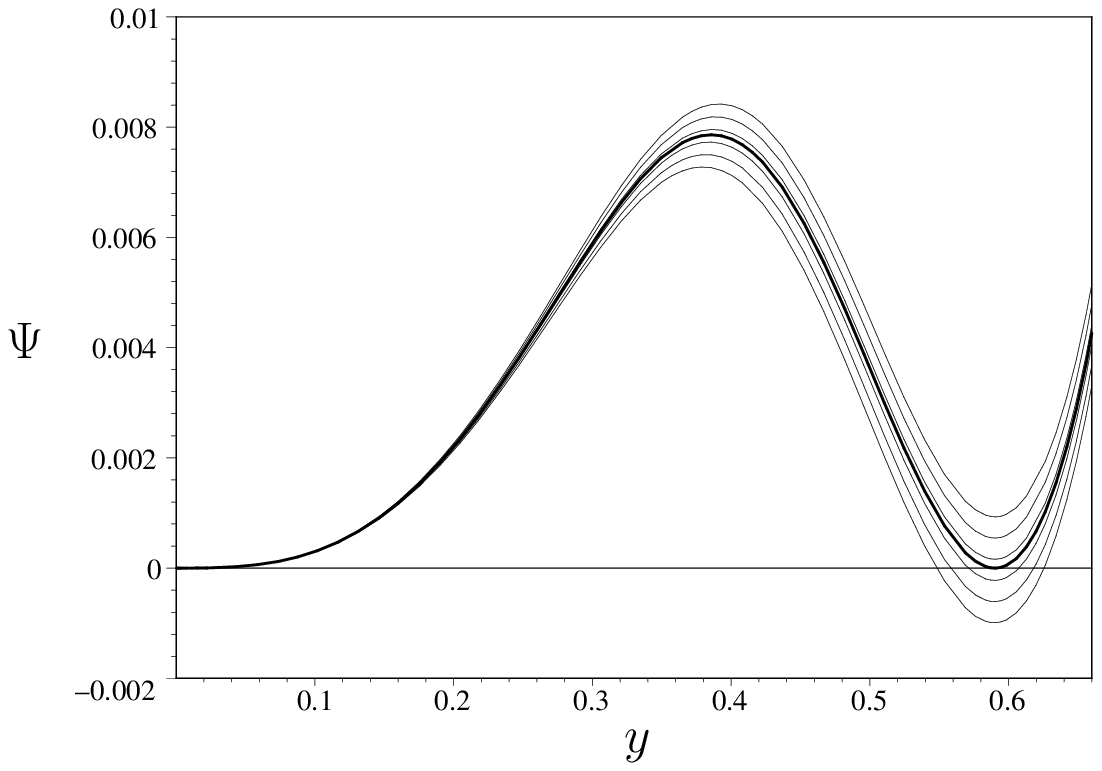,height=8cm,angle=270}
\caption{Thermodynamic potential of a single baryon in the massive GN model ($\gamma=1.2$), Eq.~(\ref{3.9}), as a function of $y\in [0,1]$. The temperature
is $T=0.03$, the $\mu$ values range from .9246 (highest curve) to .9266 (lowest curve) see also main text. A first order phase transition is found 
at $\mu_c=0.92556, y= .59$.} 
\label{fig9}
\end{center}
\end{figure}

\begin{figure}
\begin{center}
\epsfig{file=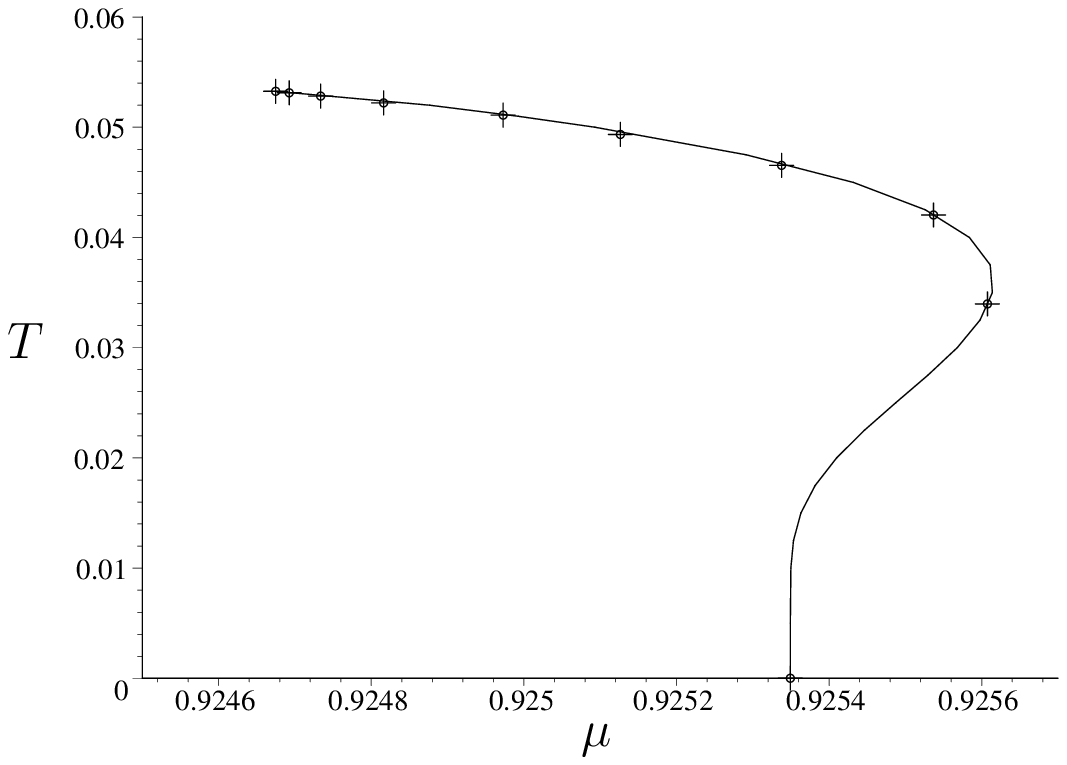,height=8cm,angle=270}
\caption{Comparison of 1st order phase boundary of a single baryon (line, this work) with 2nd order phase boundary $III$ of the massive GN model 
(crosses, Ref.~\cite{8}). The confinement parameter is $\gamma=1.2$.} 
\label{fig10}
\end{center}
\end{figure}

\begin{figure}
\begin{center}
\epsfig{file=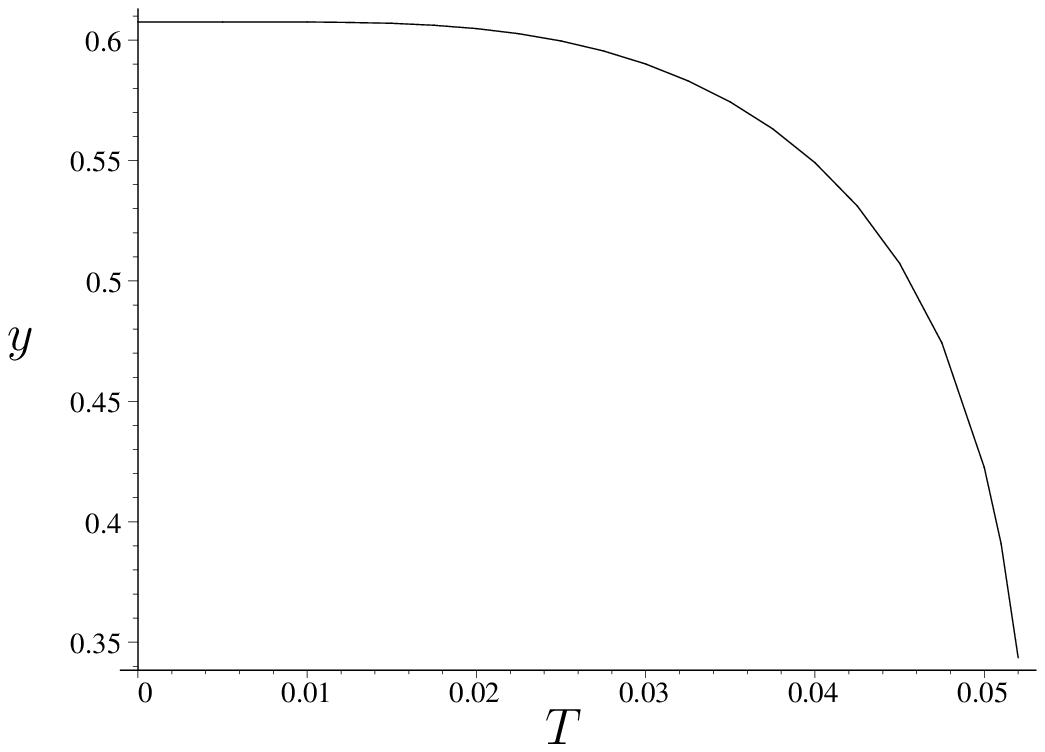,height=8cm,angle=270}
\caption{Variation of the parameter $y$ along the phase boundary of the massive GN model with $\gamma=1.2$. This is to be contrasted with the 
chiral limit where $y=1$ at all temperatures.} 
\label{fig11}
\end{center}
\end{figure}

\subsubsection{Phase transition}
\label{sec3ba}

Like in the chiral limit, we now cross the non-perturbative phase boundary $III$ on a path at constant $T$. To this end, we prescribe the confinement parameter 
$\gamma$ and use the phase boundaries of the massive model shown in Fig.~\ref{fig3}. 
By way of example, take $\gamma=1.2$, the largest value studied in Ref.~\cite{8}, $T=0.03$ and $\mu=\left\{0.9246,0.9250,\mu_c,0.9258,0.9262,0.9266\right\}$ with 
$\mu_c=0.92556795$ the critical chemical potential. In Fig.~\ref{fig9}, we have plotted the grand canonical potential for a single baryon as
a function of $y$ for all of these $\mu$-values. We observe a first order phase transition at a critical chemical potential coinciding with a point on the 
phase boundary $III$ of the massive GN model. Unlike in Fig.~\ref{fig4} at $\gamma=0$, here the minimum is not at $y=1$ but lies inside the interval [0,1].
Its position can be obtained by demanding that both $\Psi$ and $\partial_y \Psi$ vanish. Repeating this calculation for different temperatures along the phase boundary,
we obtain the solid curve shown in Fig.~\ref{fig10}. The crosses are taken from Ref.~\cite{8} and lie on the curve $III$ of the $\gamma=1.2$ plot in Fig.~\ref{fig3}.
Since we now have many more points than the previous calculation \cite{8}, we have drawn the new results as a curve and the old results as crosses,
the opposite of what we had done in Fig.~\ref{fig5}. The agreement is once again perfect. This confirms our ansatz for the mean field of the 
baryon at finite $T,\mu$ also for the massive GN model. During this calculation, we had to determine the shape parameter $y$ for each temperature by minimization. The
result is shown in Fig.~\ref{fig11}. The value at $T=0$ coincides with that of a baryon in vacuum with maximal fermion number, i.e., the solution
of Eq.~(\ref{3.3}) for the choice $n=N$ ($y=0.607533$). In the chiral limit, we had found that the $y$ parameter stays at the same value ($y=1$) for all temperatures. Here this
is not the case, but $y$ decreases monotonically by almost a factor of 2, see Fig.~\ref{fig11}.  

Once again the first order phase transition of a single baryon translates into a 2nd order transition in the infinite medium. 
Since we shall prove the agreement between full HF calculation and single baryon stability analysis analytically in the following section, it
is not necessary to repeat the whole calculation for different $\gamma$ values. 

What about self-consistency? Only by showing self-consistency can we be sure that the HF solution is correct, independently of whether the 
full solution is known. In the chiral limit, this was indeed possible analytically, see Sect.~\ref{sec2ba}. Here, we start from the self-consistency condition in the form
\begin{eqnarray}
- \frac{1}{Ng^2} (S-m_0) & = & \langle \bar{\psi}\psi \rangle_{\rm th}
\nonumber \\
& = & \bar{\psi}_0^{(+)} \psi_0^{(+)} \left( \frac{1}{e^{\beta(E_0-\mu)}+1}- \frac{1}{e^{-\beta(E_0+\mu)}+1} \right)
\nonumber \\
& & +  \int \frac{dq}{2\pi} \bar{\psi}_q^{(+)} \psi_q^{(+)} \left(  \frac{1}{e^{\beta(E_q-\mu)}+1}+ \frac{1}{e^{\beta(E_q+\mu)}+1}-1 \right).
\label{3.10}
\end{eqnarray}
We have used the fact that the scalar densities of positive and negative energy states have opposite signs.
The positive energy bound state and continuum spinors are known from the baryon problem \cite{13,17} and yield
\begin{eqnarray}
\bar{\psi}^{(+)}\psi^{(+)} & = & - \frac{\sqrt{1-y^2}}{2y} (S-M),
\nonumber \\
\bar{\psi}_q^{(+)}\psi_q^{(+)} & = &  \frac{S}{E_q} \left( 1 + \frac{(1-y^2)M^2}{q^2+y^2M^2}\right) - \frac{M}{E_q} \frac{(1-y^2)M^2}{q^2+y^2M^2} .
\label{3.11}
\end{eqnarray}
Using the gap equation (\ref{3.1}), the left hand side of (\ref{3.10}) becomes
\begin{equation}
- \frac{1}{Ng^2} (S-m_0) = - \frac{S}{\pi} \left(\ln \frac{\Lambda}{m} + \gamma\right) + \frac{\gamma m}{\pi} .
\label{3.12}
\end{equation}
Thus the self-consistency condition can be split up into one equation involving all terms proportional to $S$ and a second equation not containing $S$.
By combining the resulting two $x$-independent equations with the mass equation (\ref{3.7}), it can be reduced to a single condition,
\begin{eqnarray}
0 & = & \frac{\gamma m}{M\pi} - \frac{\sqrt{1-y^2}}{y \pi}  \arctan \left(\frac{\sqrt{1-y^2}}{y} \right) - \frac{\sqrt{1-y^2}}{2y} \left( \frac{1}{e^{\beta(E_0-\mu)}+1}- \frac{1}{e^{-\beta(E_0+\mu)}+1} \right)
\nonumber \\
& & + M^2 (1-y^2) \int \frac{dq}{2\pi} \frac{1}{E_q(q^2+y^2M^2)}\left(  \frac{1}{e^{\beta(E_q-\mu)}+1}+ \frac{1}{e^{\beta(E_q+\mu)}+1} \right).
\label{3.13}
\end{eqnarray}
We want to show that this is equivalent to minimizing $\Psi/N$ with respect to $y$. It is straightforward to differentiate all terms in Eqs.~(\ref{2.22},\ref{3.9})
except for the integral in the last line of (\ref{2.22}) which we denote as ${\cal I}$,
\begin{equation}
{\cal I} = \frac{1}{\pi \beta} \int_{0}^{\infty} dq \frac{2yM}{q^2+y^2M^2} \ln \left[ \left( 1+e^{-\beta(E_q-\mu)} \right) \left( 1+e^{-\beta(E_q+\mu)} \right) \right].
\label{3.14}
\end{equation}
If we would naively take the derivative of the integrand of ${\cal I}$ with respect to $y$, we would clearly generate structures not present
in the self-consistency condition (\ref{3.13}). The trick is to first change the integration variable $q \to yQ$, differentiate with respect to $y$ and then
undo the change of variables again, with the result
\begin{equation}
\partial_y {\cal I}  =   \int_0^{\infty} dq \left(- \frac{2M}{E_q} + \frac{2M^3y^2}{E_q(q^2+y^2M^2)} \right) \left(  \frac{1}{e^{\beta(E_q-\mu)}+1}+ \frac{1}{e^{\beta(E_q+\mu)}+1} \right) .
\label{3.15}
\end{equation}
The first term can be eliminated by using the mass equation (\ref{3.7}). The 2nd term has precisely the form of the integral appearing in the self-consistency condition
(\ref{3.13}). Performing all of these steps, one finds indeed that $\partial_y (\Psi/N)$ and the self-consistency conditions agree 
up to an irrelevant overall factor. This shows that the mean field ansatz for the baryon in the massive GN model was
correct, independently of the full solution.

\subsubsection{Thermodynamic observables}
\label{sec3bb}

\begin{figure}
\begin{center}
\epsfig{file=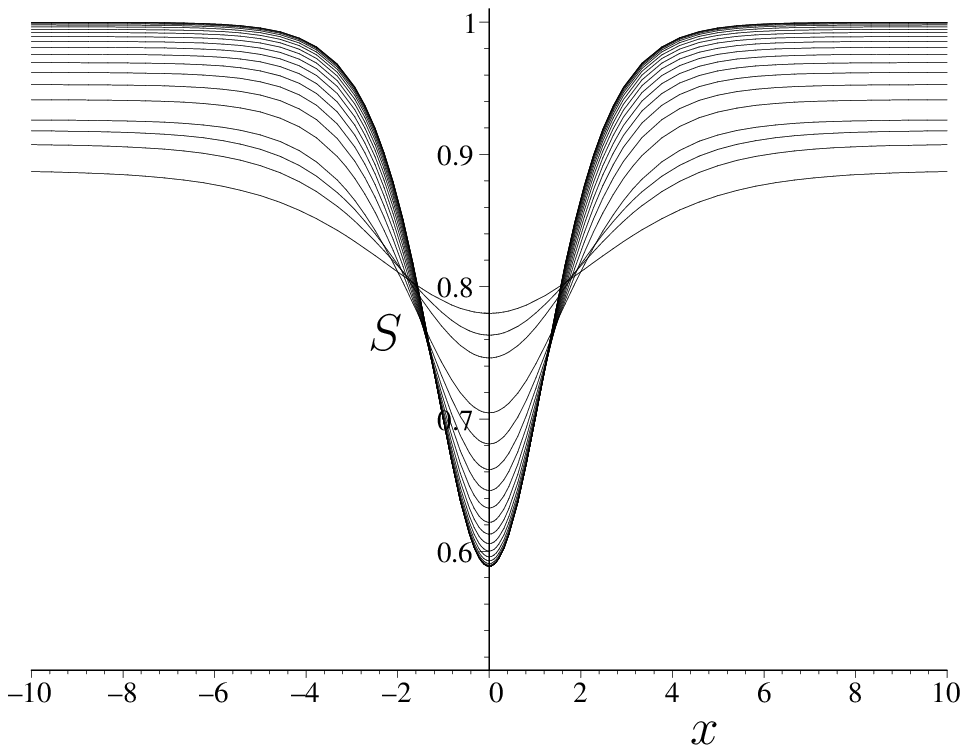,height=8cm,angle=270}
\caption{Illustration of melting of the baryon ($\gamma=1.2$) with rising temperature along the phase boundary. The steepest curve belongs to $T=0$,
the softest one to $T=.052$. The dominant factor here is the
decrease of $y$, see Fig.~\ref{fig11}, since the variation of the thermal mass is less rapid (see the circles in Fig.~\ref{fig13}). Compare to Fig.~\ref{fig6}
in the chiral limit.}
\label{fig12}
\end{center}
\end{figure}

\begin{figure}
\begin{center}
\epsfig{file=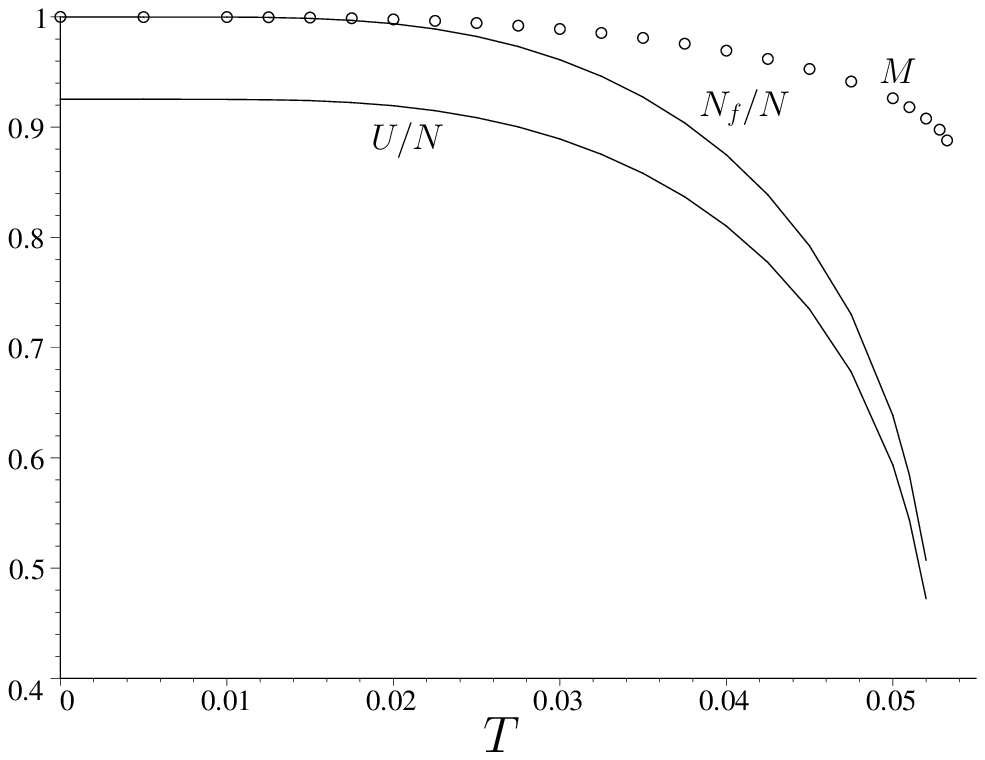,height=8cm,angle=270}
\caption{Thermodynamic observables of a single baryon ($\gamma=1.2$) along the phase boundary, plotted against temperature. Curves: 
Fermion number $N_f$ and internal energy $U$ per flavor. Circles: Thermal mass $M$. Compare to Fig.~\ref{fig7} in the chiral limit.} 
\label{fig13}
\end{center}
\end{figure}

\begin{figure}
\begin{center}
\epsfig{file=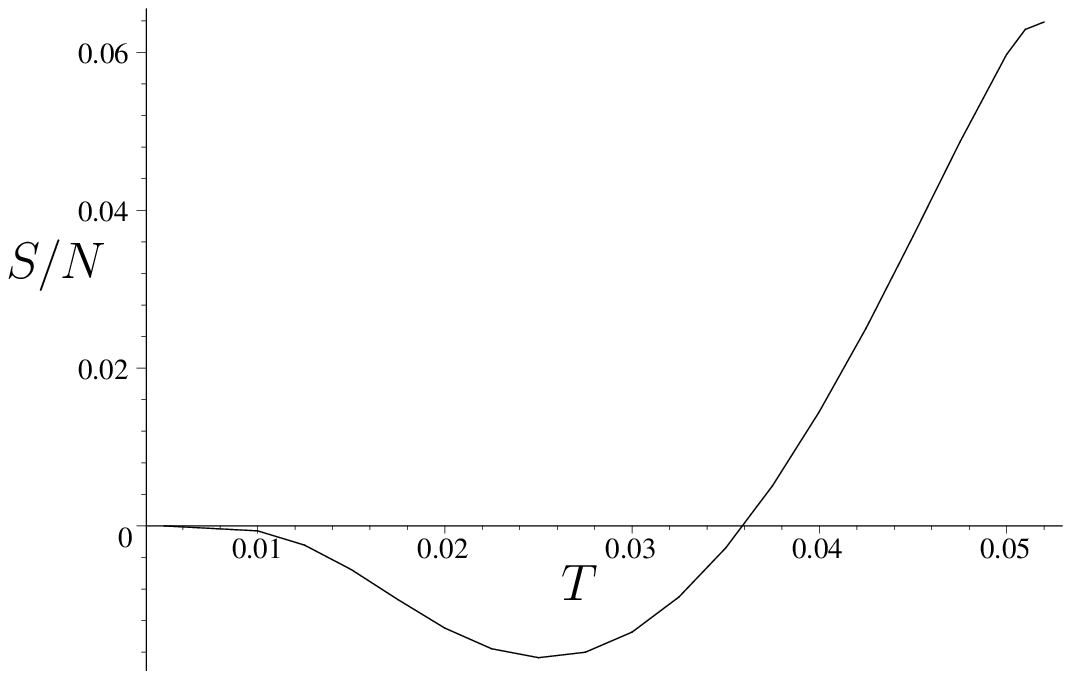,height=8cm,angle=270}
\caption{Entropy of a single baryon ($\gamma=1.2$) along the phase boundary, plotted vs. temperature. Compare to Fig.~\ref{fig8} in the chiral limit.} 
\label{fig14}
\end{center}
\end{figure}

In Fig.~\ref{fig12}, we illustrate melting of the DHN baryon in the massive GN model with increasing temperature. 
To further characterize the single baryon in thermal equilibrium, we have also computed the standard observables fermion number ($N_f$), entropy ($S$) 
and internal energy ($U$).
The results are qualitatively similar to the ones in the chiral limit, see Fig.~\ref{fig13} for $N_f$ and $U$ and Fig.~\ref{fig14} for $S$. The main difference is the fact
that the correlation between $N_f, U$ and the thermal mass $M$ is less pronounced here.

\section{Derivation from the full solution}
\label{sec4}

Since the full HF solution of the massive GN model at finite $\mu$ and $T$ is known, we have the unique chance to rederive the above results from the
exact answer for the soliton lattice in Ref.~\cite{8}. This will shed some additional light on the type of expansion involved and on its relation to the stability analysis.
We start from Eqs.~(51,53) in Ref.~\cite{8} for the thermodynamical potential of the massive GN model. As explained in that work, the non-perturbative phase
boundary corresponds to $\kappa=1$, where $\kappa$ is the elliptic modulus of all Jacobi elliptic functions and elliptic integrals appearing in the formalism.
Therefore we set
\begin{equation}
\kappa=1-8 \epsilon
\label{4.1}
\end{equation}
and expand the grand potential density around $\epsilon=0$ (the factor of 8 in (\ref{4.1}) simplifies some expressions below). It turns out that the resulting grand
canonical potential is not analytic in $\epsilon$ at $\epsilon=0$, but the leading correction goes like $1/\ln(\epsilon)$. All powers of $\epsilon$ and higher orders of $1/\ln \epsilon$ can be neglected
for our present purpose. This logarithmic dependence arises from the singularity of the complete elliptic integral ${\bf K}(\kappa)$ at $\kappa=1$. After a lengthy 
calculation, we find
\begin{equation}
\frac{\psi}{N} \approx \frac{\psi(\kappa=1)}{N} + \frac{1}{\ln \epsilon} \frac{\delta \psi}{N}.
\label{4.2}
\end{equation}
The first term agrees with the grand potential of the homogeneous phase, Eq.~(\ref{3.6}). For the correction term $\delta \psi/N$ we find
\begin{eqnarray}
\frac{\delta \psi}{N} & = & -\frac{2 y^2 M^2}{\pi} \left(1+ \ln \frac{m}{M} + \frac{\sqrt{1-y^2}}{y} \arctan \frac{\sqrt{1-y^2}}{y} \right)  - \frac{2 \gamma y M}{\pi} \left( m  {\rm \,artanh\,}y-y M\right)
\nonumber \\
& & - y M \mu +  \frac{yM}{\beta} \ln \left[ \left( 1+e^{-\beta(E_0-\mu)}  \right) \left(1 + e^{\beta(E_0+\mu)} \right) \right]
\nonumber \\
& &  - \frac{2 y^2 M^2}{\pi \beta} \int_0^{\infty} dq \frac{1}{q^2+y^2M^2} \ln \left[ \left( 1+e^{-\beta(E_q-\mu)}  \right) \left(1 + e^{-\beta(E_q+\mu)} \right) \right].
\label{4.3}
\end{eqnarray}
What type of expansion is it in terms of physics? The spatial period of the crystal lattice near $\kappa=1$ is
\begin{equation}
\ell = \frac{2 {\bf K} }{yM} \approx - \frac{\ln \epsilon}{yM},
\label{4.4}
\end{equation}
We can then re-write expansion (\ref{4.2}) in the form
\begin{equation}
\frac{\psi}{N} = \frac{\psi_{\rm hom}}{N} + \frac{1}{\ell} \frac{\Psi}{N}.
\label{4.5}
\end{equation}
This is the beginning of a power series expansion in the inverse spatial period of the crystal. The first term corresponds to infinite period and a homogeneous, massive
Fermi gas. The first order correction is proportional to $\Psi$, the grand potential of a single baryon, Eq.~(\ref{3.9}). The prefactor $1/\ell$
is just what is needed to convert the potential into a density, provided we identify $\ell$ with the size of the baryon. So, in some sense, the phase boundary $III$ is also
perturbative, not in the strength of the deformation but in its inverse size.
 
This comparison raises one last question: In the full calculation, one minimizes $\delta \psi(y)$ with respect to $y$, in the single baryon calculation one minimizes
$ \ell \delta \psi \sim y^{-1} \delta \psi(y)$. These two expressions differ by a factor of $y$. In general, this would not be equivalent. Here however, since
\begin{equation}
\frac{d}{dy}y f(y) = f(y) + y \frac{d}{dy} f(y)
\label{4.6}
\end{equation}
and $f(y)$ also vanishes along the phase boundary, this is indeed the same. This proves that the non-perturbative phase boundary can
be computed from the grand potential of a single kink-antikink baryon and shows that the present approach is correct for arbitrary values of $\gamma$.

\section{Summary and outlook}
\label{sec5}

In the phase diagram of the GN model, the soliton crystal phase is separated from the homogeneous phases by two distinct kinds of phase boundaries,
denoted by $II$ and $III$ in Figs.~\ref{fig1},\ref{fig2}. Across curves $II$, periodic modulations with finite wave number but vanishing amplitude set in. Across curves $III$,
there is a discontinuous change in amplitude at vanishing wave number, i.e., a single kink or DHN baryon pops up. Whereas curves $II$ can be predicted 
easily with a standard stability analysis, this was not the case so far with curves $III$. This is unsatisfactory, since both phase boundaries belong
to continuous, 2nd order phase transitions. We have shown that curves $III$ are also amenable to a linear stability analysis, using as expansion 
parameter not the amplitude, but the inverse spatial period (or the wave number) of the disturbance. This requires solving the thermodynamic problem of 
a single baryon in contact with a heat bath and a particle reservoir, using the grand canonical potential. This problem has turned out to be much more
tractable than the full HF solution of hot and dense matter in the GN model. The techniques are already well known from handling baryons in the vacuum,
and no prior knowledge about the detailed crystal structure is necessary. Thus, somewhat surprisingly, in both the massless and massive GN models one
can predict the complete phase diagram without knowing anything about cnoidal functions and the Lam\'e equation, the main players in the full solution.
Even if the full solution would not be known yet, the method could ``stand alone" in the sense that the self-consistency condition could be used to
check whether a particular ansatz for the disturbance in the stability analysis is correct or not.

The solution of the Dirac-HF equation in the case of 
curves $III$ is non-perturbative. Note however that even in the case of curves $II$, naive perturbation theory cannot be used because it breaks 
down at the gaps generated by periodic potentials. One has to use almost degenerate perturbation theory (ADPT \cite{20}) instead. 
The only difference between the stability analysis of curves $II$ and $III$ is the question whether one employs a periodic or a localized distortion.
In the case at hand, the effort of both types of calculation is comparable in practice.

We do not know whether an instability against a localized disturbance of a homogeneous phase occurs in other systems as well, perhaps even in higher 
dimensions. We would be surprised if this would not be the case, notably in condensed matter physics. In the case of the massless
GN model, what happens may be described as spontaneous formation of a single domain wall in an isotropic system as a function of 
chemical potential -- the fermions are needed to stabilize the kink. As mentioned above, this is unfortunately not expected to
happen in the massive chiral GN model where both the amplitude and the wave number of the disturbance are discontinuous and the corresponding
phase transition is of 1st order.

\newpage



\end{document}